\documentclass[aps, prl, superscriptaddress, reprint,aps,showpacs]{revtex4-1}

\usepackage{graphicx}
\usepackage{tabularx}
\usepackage{dcolumn}
\usepackage{bm}
\usepackage{color}
\usepackage{amsmath}
\usepackage{amssymb}

\begin{document}

\title{Square selenene and tellurene: novel group VI elemental 2D semi-Dirac materials and topological insulators}

\author{Lede Xian}
\email{ledexian03@gmail.com}
\affiliation{Nano-Bio Spectroscopy Group and ETSF,
  Universidad del Pa\'is Vasco,
  CFM CSIC-UPV/EHU,
  20018 San Sebasti\'an, Spain}

\author{Alejandro P\'{e}rez Paz }
\affiliation{Nano-Bio Spectroscopy Group and ETSF,
  Universidad del Pa\'is Vasco,
  CFM CSIC-UPV/EHU,
  20018 San Sebasti\'an, Spain}

\author{Elisabeth Bianco }
\affiliation{Department of Chemistry, Rice University, Houston, Texas 77005, United States}

\author{Pulickel M. Ajayan}
\email{ajayan@rice.edu}
\affiliation{Department of Materials Science and Nanoengineering, Rice University, Houston, Texas 77005, United States}

\author{Angel Rubio}
\email{angel.rubio@mpsd.mpg.de}
\affiliation{Nano-Bio Spectroscopy Group and ETSF,
  Universidad del Pa\'is Vasco,
  CFM CSIC-UPV/EHU,
  20018 San Sebasti\'an, Spain}
\affiliation{Max Planck Institute for the Structure and Dynamics of Matter and Center for Free-Electron Laser Science, Luruper Chaussee 149, 22761 Hamburg, Germany.}

\date{\today}

\begin{abstract}
 With first principles calculations, we predict a novel stable 2D layered structure for group VI elements Se and Te that we call square selenene and square tellurene, respectively. They have chair-like buckled structures similar to other layered materials such as silicene and germanene but with a square unit cell rather than hexagonal. This special structure gives rise to anisotropic band dispersions near the Fermi level that can be described by a generalized semi-Dirac Hamiltonian. We show that the considerably large band gap ($\sim$0.1 eV) opened by spin-orbit coupling makes square selenene and tellurene topological insulators, hosting non-trivial edge states. Therefore, square selenene and tellurene are promising materials for novel electronic and spintronic applications. Finally, we show that this new type of 2D elemental material can potentially be grown on proper substrates, such as a Au(100) surface.
\end{abstract}

\pacs{73.43.-f,73.22.-f,61.46.-w,68.65.-k}

\maketitle

The isolation of graphene in 2004 \cite{novoselov2004} opened up a new avenue in condensed matter physics: two-dimensional (2D) materials research. Following the success of graphene \cite{ferrari2015}, intensive efforts have been devoted to exploring other 2D materials \cite{miro2014}. Among them, elemental 2D materials (composed of only one element) have attracted much attention because of their simple composition and intriguing properties \cite{zhang2016}. A number of elemental 2D materials beyond graphene have been predicted and synthesized, such as silicene \cite{seymur2009,vogt2012,feng2012,fleurence2012}, germanene \cite{germanene2014}, stanene \cite{stanene2013,stanene2015}, phosphorene \cite{phosphorene2014}, and borophene \cite{borophene2015,feng2016}, with elements ranging from group III to group V. However, no studies on group VI elemental 2D materials have ever been reported. Unlike the aforementioned elements, at ambient conditions,  most of the group VI elements have 3D bulk structures composed of 1D atomic chains or 0D atomic rings with only two-fold coordination bonding. The question remains whether these elements can form 2D atomic layers as elements from groups III-V do.

During the exploration of 2D materials, some are predicted to be topological insulators (TIs), a new quantum state of matter recently discovered \cite{kane2005,bernevig2006,konig2007,hasan2010,qi2011}. TIs have different band topology than normal insulators, giving rise to non-trivial gapless surface states at the interface between TIs and normal insulators or vacuum. These nontrivial surface states are protected by time-reversal symmetry and the spin of these states is locked with their momentum, significantly reducing back-scattering. For a 2D TI, these non-trivial symmetry-protected edge states even give rise to spin-polarized conduction channels without dissipation as back-scattering is strictly prohibited \cite{bernevig2006,konig2007}. This special property makes 2D TIs extremely appealing in novel electronic and spintronic applications.

In this study, we report for the first time the theoretical prediction of a new type of 2D layered structures for group VI elements Se and Te, which we call square selenene and tellurene, following the convention of graphene and the symmetry of the unit cell. We confirm their thermal stability with \textit{ab initio} phonon calculations and molecular dynamics (MD) simulations based on density functional theory (DFT). We found that pristine square selenene and tellurene display highly anisotropic cone-shaped dispersions near the Fermi level that can be well described by a generalized semi-Dirac Hamiltonian \cite{banerjee2009,huang2015}. Interestingly, the non-vanishing momentum-dependent on-site energy term in the Hamiltonian can give rise to negative effective mass for the low-energy carriers in the conduction band. Moreover, we show that square selenene and tellurene are 2D TIs with a relatively large bulk band gap ($\sim 0.1$ eV), enabling quantum spin Hall effect to be observed at room temperature. Finally, we show that these new types of 2D elemental materials can potentially be grown on proper substrates.

\begin{figure}
\includegraphics[width=0.45\textwidth]{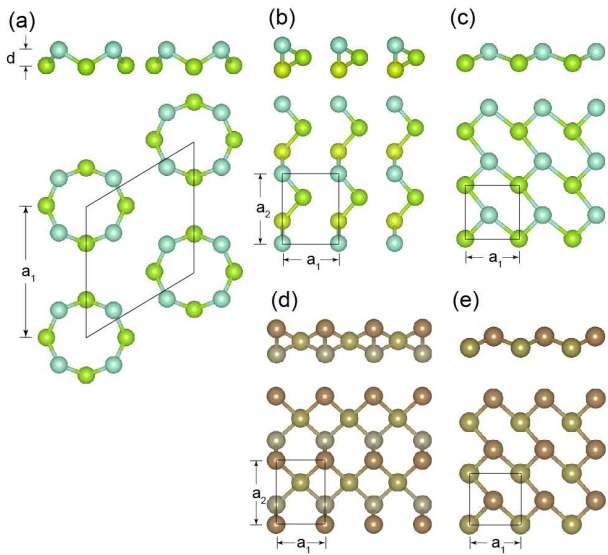}%
\caption{(color online) Lowest energy structures for freestanding Se and Te in 2D: (a) Se rings, (b) Se chains, (c) square selenene, (d) rectangular tellurene, (e) square tellurene. The side and the top views of each structure are shown in the upper and lower panels, respectively. Se and Te atoms at different heights are highlighted with different colors. The unit cell is indicated by black solid lines. The values of the structural parameters $a_1$, $a_2$, and $d$ are given in table I. \label{fig1}}
\end{figure}

 Using DFT calculations \footnote{see Supplemental Material for calculational details.\label{note1}}, we extensively investigate the lowest energy configurations for Se and Te in 2D. The main results are summarized in Fig.~\ref{fig1} and table I. As in the case of 3D, Se atoms prefer to form two-fold coordination bondings in 2D. Therefore, the lowest energy structures for Se in 2D consist of 0D atomic rings or 1D helical atomic chains, as shown in Figs.~\ref{fig1}(a) and \ref{fig1}(b), respectively. The Se-Se distances are around 2.4 {\AA} within the rings or chains and larger than 3.1 {\AA} between atoms in adjacent chains or rings. Thus, the interaction between adjacent atomic chains and rings are dominated by van der Waals interactions. These van der Waals structures in 2D can be grown on substrates, as observed in experiments when Se is deposited on metal surfaces \cite{huang1997}. However, they are easy to break upon isolation, and the electrons are mostly confined in 1D or 0D, not suitable for 2D device applications.

 Here, we propose a novel 2D atomic layered structure for Se, as shown in Fig.~\ref{fig1}(c). Atoms in this structure are arranged in a buckled square lattice with two atoms per unit cell. The atom in the center of the unit cell is tilted toward one of the corners, making the structure very similar to those of group IV 2D materials with chair-like puckering, but with the unit cell distorted into a square. Although the cohesive energy of this square structure is higher than that of the chains structure (Fig.~\ref{fig1}(b)) by 0.13 eV/atom and that of the ring structure (Fig.~\ref{fig1}(a)) by 0.11 eV/atom, all of the atoms in the layer are connected by relatively strong covalent bonds.
 \begin{table}[h!]
 \centering
 \caption{Calculated cohesive energy $E_c$ and structural parameters ($a_1$, $a_2$ and $d$) for the lowest energy freestanding structures of Se and Te in 2D shown in Fig.~\ref{fig1}. }
 \begin{tabular*}{0.45\textwidth}{@{\extracolsep{\fill}} l c c c c}
   \hline
   structure & $E_c$ (eV) & $a_1$ (\AA) & $a_2$ (\AA) & $d$ (\AA)  \\
   \hline
   \hline
   (a) Se rings & -2.76 & 8.80 & - &1.74 \\
   (b) Se chains & -2.78& 4.01&4.98 & 1.76\\
   (c) square selenene & -2.65& 3.65 &- & 0.77 \\
   (d) rectangular tellurene  & -2.57 & 4.17&5.49&2.16 \\
   (e) square tellurene & -2.51 &4.08&- &0.92 \\
   \hline
 \end{tabular*}\label{table1}
 \end{table}

 For Te, the lowest energy structure is also composed of helical chains (Fig.~\ref{fig1}(d)). But in this case, the distance between the chains becomes so close that the atoms in adjacent chains are also connected with covalent bonds, and form a 2D network, which we call rectangular tellurene. Te in a buckled square structure (Fig.~\ref{fig1}(e)), which we call square tellurene, appears as the second lowest energy structure for Te in 2D. The energy difference between the two tellurene is only 0.06 eV/atom in the freestanding form.  Such difference in cohesive energy can be easily compensated by the difference in adsorption energy when tellurene is grown on proper substrates, as we will discuss later. Because of their interesting electronic properties, we will focus on square selenene and square tellurene \footnote{see also Fig.~S3 in the Supplemental Material for their detailed atomic configurations} hereafter.

\begin{figure}
\includegraphics[width=0.48\textwidth]{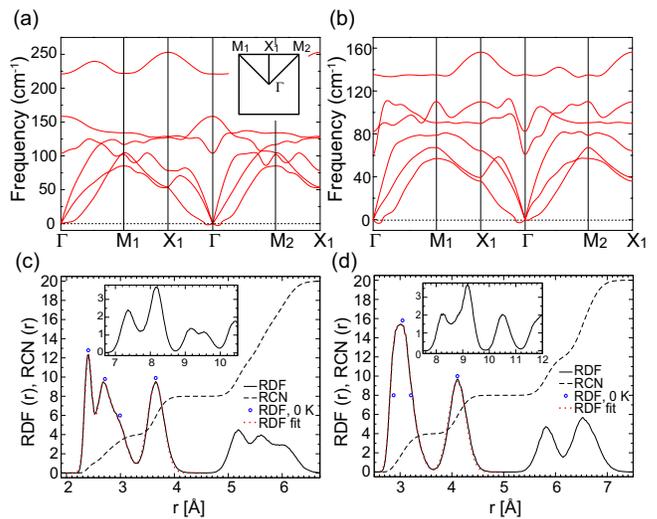}%
\caption{(color online) Top: Phonon dispersions for square selenene (a) and tellurene (b). Inset shows the first Brillouin zone of the systems. Bottom: Radial distribution function (RDF) and radial coordination number (RCN) from \textit{ab initio} MD simulations at 300 K for square selenene (c) and tellurene (d). Insets show the RDF at large distances.  \label{fig2}}
\end{figure}
 To further check the mechanical stability of square selenene and tellurene, we calculated the phonon dispersions, and the results are shown in Figs.~\ref{fig2}(a) and \ref{fig2}(b). As can be seen in these figures, there are no negative frequencies over the entire Brillouin zone, indicating that both structures are stable in the freestanding form. We also check the thermal stability of our freestanding nanostructures at room temperature using \textit{ab initio} MD simulations. During the whole MD simulations, the layered structures of square selenene and tellurene are preserved. At 300 K, our constant energy MD simulations show that the essential dynamics of selenene and tellurene monolayers are dominated by transverse harmonic modes of the membrane (see Supplemental Movie S1). The root-square mean deviation (RSMD) with respect to the initial equilibrated structure fluctuated between 0.5 and 1 {\AA}, and the Lindemann index \cite{lindemann1910} is less than $1 \% $, confirming the thermal stability of both 2D nanostructures at room temperature.

\begin{figure}
\includegraphics[width=0.45\textwidth]{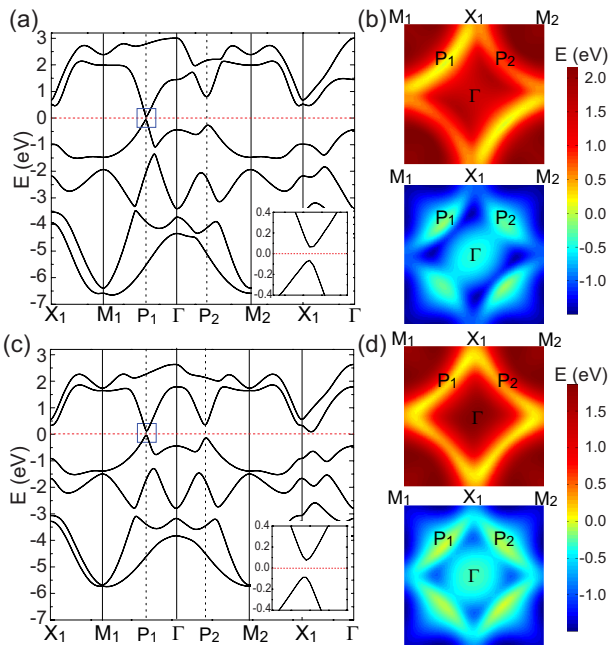}%
\caption{(color online) Top: electronic properties for square selenene: (a) Band structure. Inset: Zoom-in of the band structure in a small region around the semi-Dirac cone at $P_1$ as indicated by a blue box.  (b) Band contour for the bottom conduction band (upper panel) and the top valence band (lower panel) in the first Brillouin zone. Bottom: (c-d) corresponding results for square tellurene. SOC is included for all results. \label{fig3}}
\end{figure}

\begin{figure}
\includegraphics[width=0.42\textwidth]{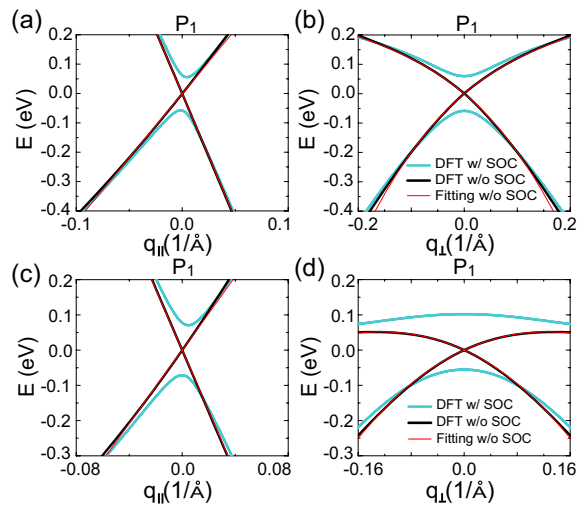}%
\caption{(color online) Top: low-energy band dispersions around P$_1$ for square selenene in the directions parallel (a) and perpendicular (b) to the $\Gamma{M_1}$ direction. Bottom: (c-d) corresponding results for square tellurene.  \label{fig6}}
\end{figure}

Radial distribution function (RDF) and radial coordination numbers (RCN) for square selenene and tellurene in the NVE ensemble (constant energy, no thermostats) are computed and shown in Figs.~\ref{fig2}(c) and \ref{fig2}(d), respectively. The RCN values give the average number of neighbors within a distance r from a given central atom. The absolute values for RDF are arbitrary because volume is not well defined for these quasi-2D buckled systems. At 300 K, the RDFs show the expected thermal broadening for all peaks, and order is preserved even at long distances (see inset) confirming the absence of amorphization. To quantify the thermal broadening, we performed a fit (red dots) of the first and second shell peaks by using a sum of 4 normalized Gaussians centered at the 0 K RDF peak positions (blue cirles). During the fits, we only varied the height and the width of each Gaussian function. The width measures the thermally-averaged flexibility of a particular bond. In the first coordination shell, the shortest bond lengths (2.40 and 2.87 {\AA} for Se and Te systems, respectively) are the ones that exhibit less thermal fluctuations in the RDF at 300 K, with half of the Gaussian width (0.08 \AA) than the largest bonds (0.17 {\AA}, peaks at 3.65 and 4.11 {\AA} for Se and Te systems, respectively). Overall, the agreement between 300 K RDF's and its respective fit is excellent indicating that the dynamics are in the harmonic regime.

 Having demonstrated the mechanical stability of square selenene and tellurene, next we investigate their electronic properties. The calculated band structures including spin-orbit coupling (SOC) are shown in Figs.~\ref{fig3}(a) and \ref{fig3}(c). Interestingly, there exist Dirac-cone-like dispersions  at $P_1$ (along the $\Gamma{M_1}$ direction) in the Brillouin zone for both systems (indicated by a blue box in Figs.~\ref{fig3}(a) and \ref{fig3}(c)). But unlike the Dirac cone in group IV 2D materials, these band dispersions are highly anisotropic (see regions around $P_1$ in the band contours shown in Figs.~\ref{fig3}(b) and \ref{fig3}(d)). In fact, when SOC is turned off in the DFT calculations, these anisotropic bands display semi-Dirac dispersions: they are linear in the the $\Gamma{M_1}$ direction (black solid lines in Figs.~\ref{fig6}(a) and \ref{fig6}(c)) and parabolic-like in the perpendicular direction (black solid lines in Figs.~\ref{fig6}(b) and \ref{fig6}(d)). These band dispersions near the Fermi level are well described by a $2\times2$ generalized semi-Dirac effective Hamiltonian \cite{huang2015}:
 \begin{equation}\label{eqn1}
 \mathbf{H}(\mathbf{q})=\epsilon(\mathbf{q})\mathbb{I}_{2\times2}+\mathbf{h}(\mathbf{q})\cdot\overrightarrow{\sigma},
 \end{equation}
 where $\mathbf{q}=(q_{\parallel},q_{\perp})$ is the momentum displaced from point P$_1$ with $q_{\parallel}$ and $q_{\perp}$ defined in the directions parallel and perpendicular to the $\Gamma{M_1}$ direction, respectively, $\mathbb{I}_{2\times2}$ is the unit matrix, $\overrightarrow{\sigma}=(\sigma_x,\sigma_y,\sigma_z)$ are the Pauli matrices, $\epsilon(\mathbf{q})=-Aq_{\parallel}-Bq_{\perp}^2$ is the effective on-site energy, and
 $\mathbf{h}(\mathbf{q})=(Cq_{\parallel}+Dq_{\perp}^2,Fq_{\perp}+Gq_{\parallel}q_{\perp},0)$. The existence of the momentum-dependent on-site energy term $\epsilon(\mathbf{q})$ breaks the electron-hole symmetry of the systems. The resulting energy dispersion has the form:
 \begin{equation}\label{eqn2}
 E=-Aq_{\parallel}-Bq_{\perp}^2\pm \sqrt{(Cq_{\parallel}+Dq_{\perp}^2)^2+(Fq_{\perp}+Gq_{\parallel}q_{\perp})^2},
 \end{equation}
 matching perfectly with the DFT results (see red lines in Fig.~\ref{fig6}) \footnote{see Supplemental Table S1 for fitting parameters}. In the $\Gamma{M_1}$ direction with $q_{\perp}=0$, it reduces to $E=-Aq_{\parallel}\pm |Cq_{\parallel}|$, in which the linear dispersions are preserved. In the perpendicular direction with $q_{\parallel}=0$, the energy dispersion reduces to $E=-Bq_{\perp}^2\pm \sqrt{D^2q_{\perp}^4+F^2q_{\perp}^2}$. The momentum-dependent on-site energy term $-Bq_{\perp}^2$ even gives rise to a negative curvature in the bottom conduction band in the energy range of $0\sim0.2$ eV for square selenene and $0\sim0.05$ eV for square tellurene (see Figs.~\ref{fig6}(b) and \ref{fig6}(d)), which corresponds to negative effective mass for the charge carriers.

\begin{figure}
\includegraphics[width=0.45\textwidth]{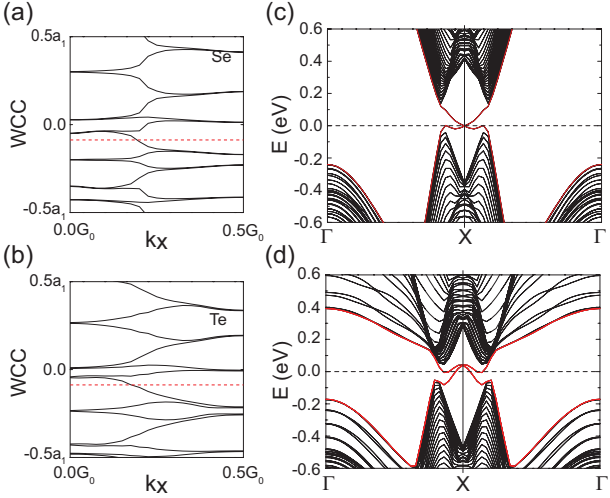}%
\caption{(color online) Evolution of the Wannier charge centers (WCC) along $k_x$ for square selenene (a) and tellurene (b). $a_1$ is the lattice constant and $G_0=2\pi/a_1$. The red dash reference line is only crossed once in both figures, indicating that the $Z_2$ invariant is 1 (non-trivial) for both systems. (c-d) band structures including SOC near Fermi level for a wide ribbon of square selenene (c) and tellurene (d). The helical edge states are highlighted with red solid lines.  \label{fig4}}
\end{figure}

  Due to the SOC, a considerable band gap $E_g$ is opened at P$_1$ (see blue solid lines in Fig.~\ref{fig6}) with $E_g=0.13$ eV for square selenene and $E_g=0.16$ eV for square tellurene. The sizes of these band gaps are comparable to that of stanene \cite{stanene2013}. This effect can also be considered in Eq.~\ref{eqn1} with the addition of a mass term $m_z\sigma_z$. For square tellurene, the SOC band gap turns the band crossing point in P$_1$ into a saddle point. Therefore, square tellurene becomes an indirect band gap semiconductor (indirect gap $E_g^I=0.14$ eV in DFT-GGA calculations) with the bottom of the conduction bands located between $X_1$ and $\Gamma$ point. For square selenene, the bottom of the conduction band remains at P$_1$ and it is a direct band gap semiconductor when the SOC is included. The additional mass term $m_z\sigma_z$ only affects the dispersion very close to P$_1$, and the effective mass for charge carriers in the range of $0.08\sim0.2$ eV remains negative in the direction perpendicular to the $\Gamma{M_1}$ direction. Such special dispersion is expected to provide an interesting platform to explore new physics and novel phenomena. The features in the band structures along $\Gamma{M_2}$ are similar to those along $\Gamma{M_1}$, but the band gap at the Fermi level at $P_2$ is opened by structural distortion, as the atom in the center of the unit cell is shifted off center along one diagonal direction (see Figs.~\ref{fig1}(c) and \ref{fig1}(e)).

 Both square selenene and tellurene have non-trivial topological properties. Here, we calculate their $\mathbf{Z}_2$ invariant from the evolution of the Wannier charge centers (WCC) during an effective adiabatic pumping process \cite{soluyanov2011,yu2011}. The results are shown in Figs.~\ref{fig4}(a) and \ref{fig4}(b), respectively. From the figures, we can see that the WCC bands (black lines) and the reference line (red dash line) only cross once for both systems. Therefore, their $\mathbf{Z}_2$ is 1, indicating that both systems are TIs. One of the most important features of TIs is that they host gapless helical edge states with spin and momentum locked. This is confirmed by our DFT calculations for wide ribbons of selenene and tellurene. Figs.~\ref{fig4}(c) and \ref{fig4}(d) show the band structures for zigzag ribbons of selenene and tellurene, with the width being 12.4 nm and 13.9 nm, respectively (see Supplemental Fig.~S6). We can clearly see the gapless edge bands (red solid lines) lie in between the bulk band gaps for both systems, confirming their nontrivial topological properties. This is distinctly different from their bulk allotropes in 3D, for which considerable amount of shear strain is required to transform them from trivial semiconductors to TIs \cite{agapito2013}.

 Earlier we noted that square selenene and tellurene do not have the lowest cohesive energy among all the 2D structures for Se and Te, respectively. However, when all of them are placed on top of a substrate, the energy ordering between different structures may change due to substrate effects. Although structures composed of helical chains have the lowest cohesive energy in the gas phase, they tend to obtain less adsorption energy as they are more buckled. On the other hand, with less height variation, square selenene and tellurene may become the most energetically favorable structures on substrates with square symmetry and matching lattice constants. For demonstration, we calculated the formation energy of square tellurene and Te helical chains (rectangular tellurene, see Fig.~\ref{fig1}(d)) on a Au(100) surface. We found that square tellurene has a formation energy 0.02 eV/atom lower than that of rectangular tellurene, indicating that square tellurene is more energetically favorable to be grown on a Au(100) surface. More interestingly, when we calculate the atomic structures of Te deposited on a Au(100) surface from 0.5 monolayer (ML) to 2 ML (see Fig.~S7 in the Supplemental Material), we found that square tellurene is grown layer by layer with the same square pattern on the top surface. This matches nicely with the experimental observation of Ikemiya \textit{et al.} who reported a $\sqrt{2} \times \sqrt{2}$ pattern for Te deposition on Au(100) surface with coverages up to 5 ML using high resolution atomic force microscopy (HR-AFM) \cite{ikemiya1996}. A similar situation applies to square selenene: proper substrates satisfying square symmetry and lattice matching conditions are expected to stabilize square selenene instead of other 2D Se structures. 

 The novel structures we predict here provide realization of TIs in a square lattice with lower symmetry than those discussed by Slager \textit{et al.} for a square lattice in the space group classification of topological band-insulators \cite{slager2013}. These structures also go beyond the simple model discussed by Young \textit{et al.} for Dirac semimetals in 2D \cite{young2015}, because their model is limited to s state electrons, while ours are dominated by p-state hybridizations around the Fermi level. Thus, the special structures predicted here provide a new template for the exploration of other novel Dirac or semi-Dirac semimetals and TIs in 2D \cite{ren2016}.

 In conclusion, we report here for the first time novel 2D structures for Se and Te, namely, square selenene and tellurene. We confirm their stability with \textit{ab initio} phonon calculations and MD simulations. We found they have very interesting electronic structures with two gapped semi-Dirac cones in the square Brillouin zone, and they are 2D TIs with non-trivial topological properties. These intriguing properties make them interesting platforms for exploring novel physics and unusual phenomena stemming from the semi-Dirac dispersions and promising materials for 2D electronic and spintronic applications. Finally, using Te on a Au(100) surface as an example, we demonstrate that they can be grown on proper substrates with square symmetry.

 We acknowledge financial support from the European Research Council (ERC-2015-AdG-694097), Spanish grant (FIS2013-46159-C3-1-P), Grupos Consolidados (IT578-13), AFOSR Grant No. FA2386-15-1-0006 AOARD 144088, H2020-NMP-2014 project MOSTOPHOS (GA No. 646259) and COST Action MP1306 (EUSpec). A.P.P. acknowledges postdoctoral fellowship from the Spanish ``Juan de la Cierva-incorporaci\'{o}n" program (IJCI-2014-20147). E.~B. acknowledge the support by the National Science Foundation Graduate Research Fellowship under Grant No. (DGE-1450681).

\end{document}